\documentclass[superscriptaddress,nofootinbib,nobibnotes,amsmath,amssymb,aip,preprint]{revtex4-1}
\usepackage{graphicx}
\usepackage{dcolumn}
\usepackage{bm}
\usepackage[T1]{fontenc}
\usepackage{mathptmx}
\usepackage{color}
\usepackage{natbib}

\begin{document}

\title{Polarization amplification by spin-doping in nanomagnetic/graphene hybrid systems}

\author{Tim Anlauf}
\affiliation{Center for Hybrid Nanostructures, Universit\"at Hamburg, Luruper Chaussee 149, 22761 Hamburg}
\author{Marta Prada}
\affiliation{I. Institute for Theoretical Physics, Universit\"at  Hamburg HARBOR, Geb. 610 Luruper Chaussee 149, 22761 Hamburg}
\author{Stefan Freercks}
\affiliation{Center for Hybrid Nanostructures, Universit\"at Hamburg, Luruper Chaussee 149, 22761 Hamburg}
\author{Bojan Bosnjak}
\affiliation{Center for Hybrid Nanostructures, Universit\"at Hamburg, Luruper Chaussee 149, 22761 Hamburg}
\author{Robert Fr\"omter}
\affiliation{Center for Hybrid Nanostructures, Universit\"at Hamburg, Luruper Chaussee 149, 22761 Hamburg}
\author{Jonas Sichau}
\affiliation{Center for Hybrid Nanostructures, Universit\"at Hamburg, Luruper Chaussee 149, 22761 Hamburg}
\author{Hans Peter Oepen}
\affiliation{Center for Hybrid Nanostructures, Universit\"at Hamburg, Luruper Chaussee 149, 22761 Hamburg}
\author{Lars Tiemann}
\email[]{Correspondence: lars.tiemann@physik.uni-hamburg.de}
\affiliation{Center for Hybrid Nanostructures, Universit\"at Hamburg, Luruper Chaussee 149, 22761 Hamburg}
\author{Robert H. Blick}
\affiliation{Center for Hybrid Nanostructures, Universit\"at Hamburg, Luruper Chaussee 149, 22761 Hamburg}

\date{\today}
\newpage

\begin{abstract}
The generation of non-equilibrium electron spin polarization, spin transport, and spin detection are fundamental in many quantum devices. We demonstrate that a lattice of magnetic nanodots enhances the electron spin polarization in monolayer graphene via carrier exchange. We probed the spin polarization through a resistively-detected variant of electron spin resonance (ESR) and observed resonance amplification mediated by the presence of the nanodots. 
Each nanodot locally injects a surplus of spin-polarized carriers into the graphene, and the ensemble of all  'spin hot spots' generates a  non-equilibrium electron spin polarization in the graphene layer at macroscopic lengths. 
This occurs whenever the interdot distance is comparable or smaller than the spin diffusion length. 
\end{abstract}

\maketitle
\noindent  

\section{INTRODUCTION}

The lifetime of a spin information in a two-dimensional carrier system can vary considerably between a few picoseconds and several microseconds, depending on the host material, imposed confinements, the temperature and various types of interactions \cite{Fabian1999}. Carrier transport with long spin lifetimes, or equivalently with long relaxation lengths, is key to store, transport, and compute quantum information using the electron spin. The low intrinsic spin-orbit interaction in monolayer graphene, the low atomic number of its carbon atoms and a very small hyperfine coupling between the nuclear and electron spins theoretically permits such long spin lifetimes in the range of microseconds \cite{Hernando2006, Dora2010, Jeong2011, Dugaev2011, Pesin2012, Han2014}. In real graphene devices, however, the lifetime is limited by the interaction with the substrate and/or adatoms, which contribute extrinsic spin-orbit coupling and enhance parasitic spin relaxation mechanisms \cite{Ertler2009}. Experimentally determined spin lifetimes are thus found to be two orders of magnitudes smaller than theoretically anticipated \cite{Tombros2007, Popinciuc2009,Pi2010, Han2010, Han2011, Yang2011, Guimaraes2012, Kozikov2012, Lundeberg2013, Drogeler2014, Guimaraes2014, Drogeler2016, Singh2016}. In devices fabricated from graphene synthesized by chemical vapor deposition (CVD) on a SiO$_2$ substrate, we are confronted by a multitude of intrinsic and extrinsic defects that promote spin flips and govern spin relaxation. 

In this work, we address the question whether CVD graphene can be refined to still maintain a high (non-equilibrium) spin polarization over macroscopic distances. Here, we decorated the supporting substrate with a lithographically defined hexagonal lattice of magnetic nanodots, with an average lattice constant comparable to the spin diffusion length, $\lambda_\mathrm{S}$, of the electrons in the graphene layer. In electron spin resonance (ESR) experiments under a magnetic field perpendicular to the sample, and at low temperatures, we observe that graphene in this nanomagnet hybrid device experiences an enhancement of the electron spin polarization. We verify the enhanced polarization by analyzing the resonance peaks and comparing them to a reference sample without nanodots. The magnetic nanodots appear to act as spin hot spots (or spin dopants) that boost the spin polarization in the nearby 2D carrier system. These nanomagnetic/graphene hybrid structures are not classical spin valve devices, however, they constitute an alternative strategy for controlling the spin polarization and spin transport in two-dimensional materials with strong spin diffusion.

\section{POLARIZATION MODEL\label{sec:model}}

We consider first an unintentionally doped graphene Hall bar structure. A magnetic field $\bm B$ splits the spin-up and -down electron states by the Zeeman energy, $E_\mathrm{Z} = g\mu_B B$, where $g\approx$1.95 \cite{Mani2012, Sichau2019} is the electron $g$-factor in graphene and $\mu_B$ the Bohr magneton. As dictated by the Fermi-Dirac statistics, a small spin  polarization is thermally induced, as the occupation of the lower spin-band $N_-$ is slightly larger than that of the upper spin-band, $N_+$ (see Supplementary Material). We obtain $P(N_+) - P(N_-) \simeq  10^{-2}-10^{-3}$ for a typical field of 1 Tesla as the Fermi level is changed by a floating gate. Although this appears to be a very low number, it induces a detectable resistivity signal. We note that also medical magnetic resonance imaging (MRI) at room temperature only operates on a tiny imbalance between the population of (nuclear) spin states in the human body.

In our resistively-detected variant of  ESR experiment at low temperatures, the spin split bands are coupled by a microwave excitation \cite{Mani2012, Sichau2019, Singh2020,Lyon2017PRL}. Whenever the microwave energy matches the Zeeman splitting, $h\nu = g\mu_B B$, the resonance condition is met, and a spin-flip is accompanied by the absorption of $h\nu$, that is, the carriers make a transition to the upper Zeeman level. A signal in the differential resistance is then observed, $\Delta \rho_{xx}(\nu) = \rho_{xx}^{\rm dark} - \rho_{xx}^{\nu}$, where $\rho_{xx}^{\rm dark}$ ($\rho_{xx}^{\nu}$) is the longitudinal resistivity in the absence (presence) of radiation. In the resonant condition, the band population increases, reducing $\rho_{xx}^{\nu}$ and consequently inducing a peak in $\Delta \rho_{xx}$. In graphene in the electron regime, the observed signal is thus proportional to the occupation probability of the lower band times the probability to find a vacancy in the upper band, that is, to $f(\varepsilon_-)\cdot[1-f(\varepsilon_+)]$, where $f$ is the Fermi-Dirac distribution function and $\varepsilon_\pm =\mu_F\pm E_\mathrm{Z}$, and the chemical potential $\mu_F$ being a function of the carrier density, $n$. In order to obtain the functional, we employ a mean-field approximation (Supplementary Material of Ref. \onlinecite{Sichau2019}), namely, density-functional theory (DFT) with a local electron-electron interaction and obtain a linear relation, $\mu_F \simeq \gamma n$, with $\gamma = 1.2 \times 10^{-11}$ meV$\cdot$cm$^2$. We then employ a one-parameter fit to the signal, namely, 
\begin{equation}
\Sigma(n) = \alpha_s f(\varepsilon_-)\cdot[1-f(\varepsilon_+)],
\label{eqSigma}
\end{equation}
where $\alpha_s$ is found empirically. Note that the signal decreases exponentially with $E_F$, as both bands become fully occupied.  

\begin{figure}[!]
   \includegraphics[width=0.25\textwidth, angle=0]{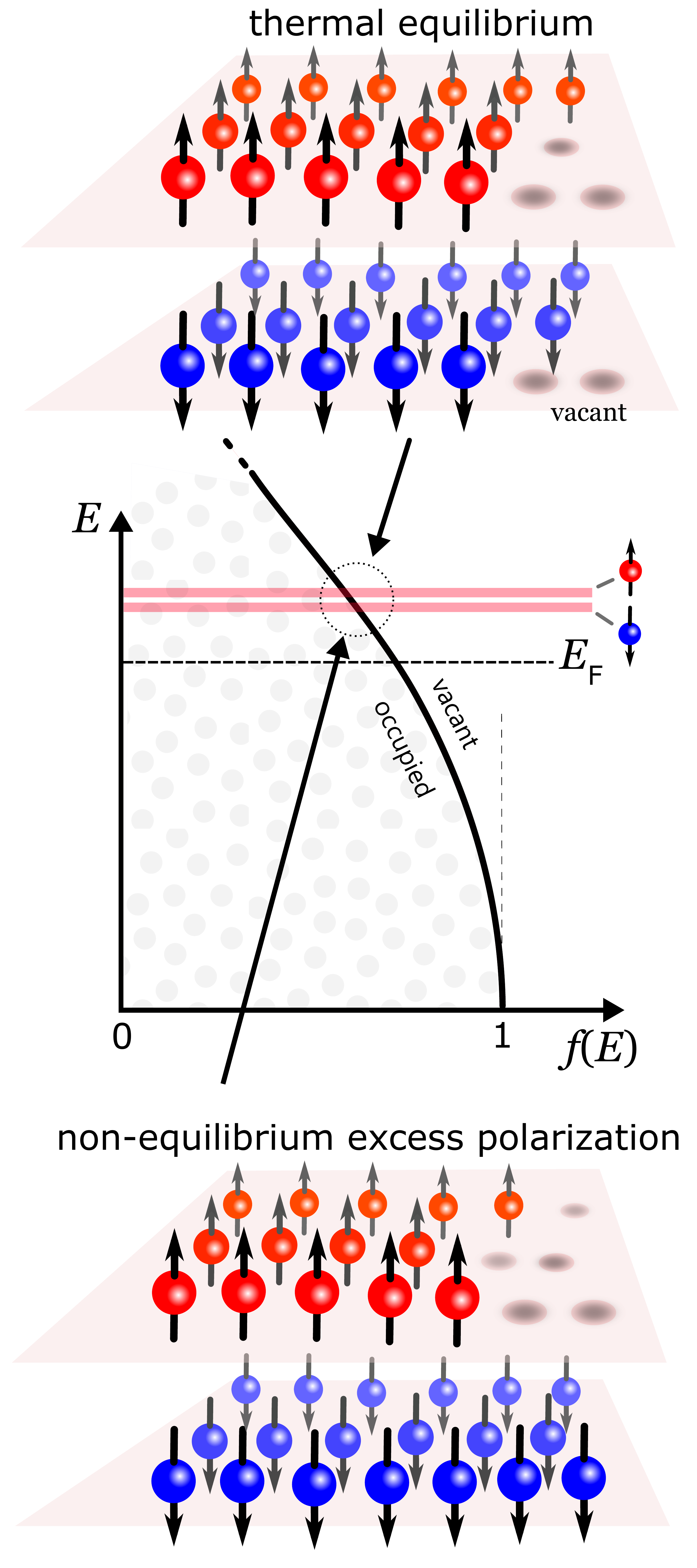}
   \caption{Schematic illustration of exemplary spin populations at a finite magnetic field. The center panel shows a section of the Fermi distribution function in comparison to the upper and lower spin bands (red bars). The Fermi energy, $E_F$, dwarves the Zeeman splitting that is of the order of 100 $\mu$eV. The red layers above and below are magnifications of the spin bands and their occupations in a system without excess spin polarized carriers (top) and in the presence of additional spin sources (bottom). The resistive detection of ESR is based the small difference in the occupation of the spin down and spin up levels which is different in the top and bottom panels.}
  \label{fig1}
\end{figure}

We consider next a graphene sample on a ferromagnetic nanodot superlattice that provides spin-polarized electrons.
When a fraction of the dopants are spin-polarized, both  factors of Eq. (\ref{eqSigma}), $f(\varepsilon_-)$ and $[1-f(\varepsilon_+)]$, become enhanced, resulting in an enlarged signal by up to several orders of magnitude. Exemplary occupations of the upper and lower spin bands without and with excess spin polarized carriers are illustrated in Fig. \ref{fig1}. Our model does not assume or require a specific geometry as source for the spin polarized carriers as it is based on statistics. In the following experimental study, we exploited magnetic nanodots that boost the polarization via exchange of spin polarized carriers, to a degree that the signal is detected directly in the resistivity $\rho_{xx}^\nu$. We use this polarization model to interpret the experimental study of the spin polarization in two graphene samples.

\section{SAMPLES AND EXPERIMENTAL METHODOLOGY}

Sample A serves as a reference and consists of a Hall bar of CVD monolayer graphene on a SiO$_2$/$p$-Si substrate. The Hall bar has a width of $W$ = 22 $\mu$m and a length of 200 $\mu$m, the lateral voltage probe separation is $L$ = 100 $\mu$m. The details of the graphene processing can be found in Ref. \onlinecite{Lyon2017APL}. The graphene structure is unintentionally $n$-doped, with an intrinsic electron density of $n\simeq$ 2 $\times$ 10$^{11}$ cm$^{-2}$, which can be changed by the $p$-doped substrate that acts as a gate. 

\begin{figure}[!]
   \includegraphics[width=0.3\textwidth, angle=0]{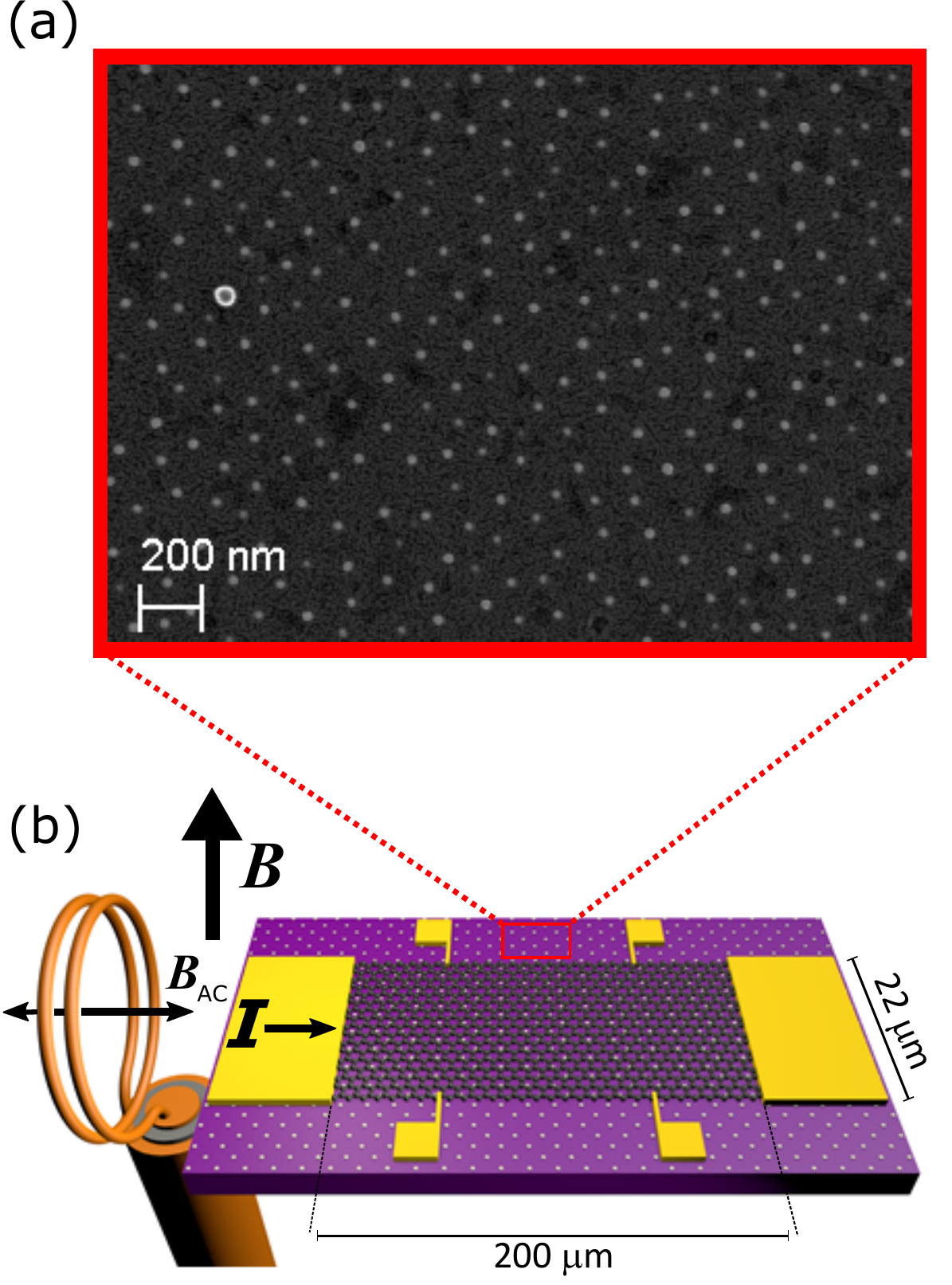}
   \caption{Sample B. (a) SEM image of the substrate with prepatterned Pt/Co nanodots. (b) Schematic illustration of the graphene Hall bar sample B on the substrate with nanodots (white dots) and the relative orientations of the microwave field generated by a nearby coil, the external magnetic field ($B$), and the low frequency transport current ($I$). Sample A has the same dimensions but rests on a SiO$_2$ without nanodots (not shown here).}
  \label{fig2}
\end{figure}

Sample B consists of the same CVD graphene material and Hall bar geometry, however, it rests on a SiO$_2$ substrate with cylindrical magnetic Pt/Co nanodots. Figure \ref{fig2}(a) shows a scanning electron microscopic (SEM) image of the nanodots on the SiO$_2$ substrate (sample B). This lattice of magnetic nanodots was created in a multistep process which is detailed in Ref. \onlinecite{Stillrich2008}. We deposited a 3 nm Pt/0.7 nm Co/7 nm Pt film on top of a SiO$_2$/$p$-Si substrate and coated the surface with SiO$_2$-filled diblock copolymer micelles. These core-shell particles self-assemble into a densely-packed, low-order hexagonal monolayer. In a O$_2$ plasma, the copolymer shells are removed, leaving behind the SiO$_2$ cores at a distance equal to twice their former shell thickness. The SiO$_2$ cores act as shadow masks during sputter milling, resulting in an array of cylindrical Pt/Co nanodots with an average diameter of 27 nm and an average height of 10 nm. The nanodots cover 1.4\% of the substrate surface with a mean separation of approximately 200 nm. We chose this separation [through the shell thickness of the micelles] because it is smaller than the typical spin diffusion lengths in our CVD graphene devices as shown in the following section. Each nanodot represents a single ferromagnetic domain with its magnetization pointing out of plane \cite{Stillrich2008, Neumann2012} due to interface anisotropy. We note that the magnetic stray field perpendicular to the nanodot surface reaches approximately 59 mT, however, at lateral distances of 200 nm the stray field is too small to (anti)ferromagnetically couple neighboring dots \cite{Neumann2013, Neumann2014}. Hence, each dot can be considered as an independent nanomagnet in contact to the graphene layer covering it.

Both samples were cooled down to nominally 1.3 K in the same custom-made vacuum probe station that is submerged in a liquid helium variable temperature insert (VTI). The cryostat is equipped with a superconducting magnet that generates a magnetic field perpendicular to the sample plane. We employ a standard lock-in method that passes a low frequency alternating current of $I$ = 2 nA and 37 Hz through the Hall bar. The lock-in amplifiers detect the longitudinal voltage drop, $V_{xx}$, and the Hall voltage, $V_{xy}$. The resulting longitudinal and Hall resistivities are calculated as $\rho_{xx}=\frac{W}{L}\cdot \frac{V_{xx}}{I}$ and $\rho_{xy}=R_{xy}=\frac{V_{xy}}{I}$, respectively. At low temperatures  and at the charge neutrality point, sample A has an (electron) density of $n\approx$ 2 $\times$ 10$^{11}$ cm$^{-2}$, a mobility of $\mu\approx$ 3200 cm$^2\cdot(\mathrm{V}\cdot\mathrm{s})^{-1}$ and a carrier mean free path for ballistic transport of $l_\mathrm{e} = \frac{h\mu}{2e}\sqrt{\frac{n}{\pi}} \approx$ 16.7 nm. Sample B is characterized by an intrinsic hole density of $p\approx$ 1.5 $\times$ 10$^{13}$ cm$^{-2}$, a mobility of $\mu\approx$ 70 cm$^2\cdot$(V$\cdot$ s)$^{-1}$ and $l_\mathrm{e} \approx$ 3.2 nm. The samples were exposed to microwaves through a two-turn Hertzian coil located next to the sample. The coil is connected through a semi-rigid coaxial wire to a frequency generator. Microwaves are applied as continuous wave (CW) with constant frequency $\nu$ and constant power amplitude, and the ESR signal is encountered by sweeping the magnetic field. The sample temperature under microwave radiation may increase up to 20-30 K.

\section{RESULTS AND DISCUSSION}
\subsection{Spin polarization in unaltered CVD graphene}

\begin{figure*}[!t]
   \includegraphics[width=0.75\textwidth, angle=0]{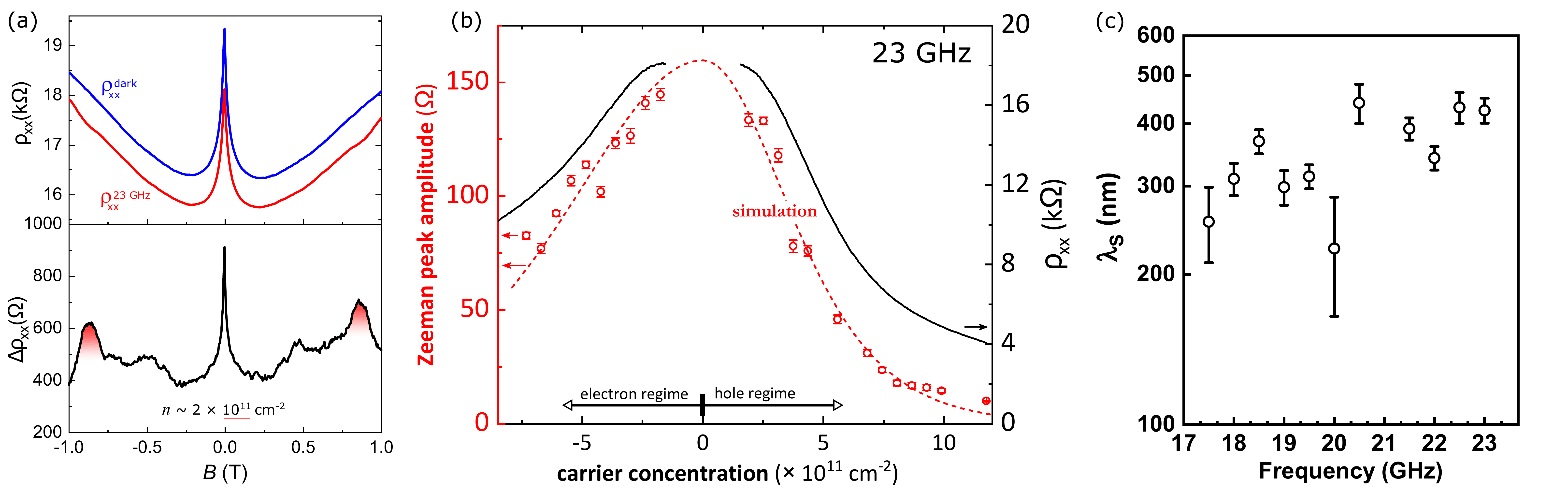} 
   \caption{Reference sample A (graphene on plain SiO$_2$). Upper panel (a): $\rho_{xx}^{\mathrm{dark}}$ measured without microwaves (blue solid line) and $\rho_{xx}^{\mathrm{23~GHz}}$ under continuous radiation of 23 GHz/21 dBm (red solid line). The back gate was tuned to the smallest density at the charge neutrality point of approximately $n \approx 2 \times$ 10$^{11}$ cm$^{-2}$. Thermal activation leads to an overall increase of the conductivity under microwave radiation. Lower panel (a): $\Delta\rho_{xx} = \rho_{xx}^{\mathrm{23~GHz}} - \rho_{xx}^{\mathrm{dark}}$. The peak centered at $B$ = 0 T is resulting from thermal activation in the weak localization regime; the  highlighted areas indicate electron spin resonances (see main text). (b) Open circles represent resonance amplitudes in ohms, taken from Lorentzian fits of the resonances that occur around 0.85 T as function of the carrier concentration. The dashed red line shows the estimated resonance amplitudes based on the model in section \ref{sec:model}. The black solid line represents the CNP in $\rho_{xx}$ measured at $B$ = 0 T (right-hand axis). A discontinuity exists because the density can not be tuned to be zero for $T >$ 0 K. (c) Spin diffusion length ($\lambda_\mathrm{S}$) versus frequency measured at a density of 2 $\times$ 10$^{11}$ cm$^{-2}$.} 
  \label{fig3}
\end{figure*}

In plain graphene samples on an insulating substrate, the resonance peaks are usually buried in the resistive background. The issue lies in the effective temperature increase under microwave radiation that enhances the conductivity of the graphene in the whole magnetic field range. The signal from the carriers in resonance is only observed in the differential resistivity\cite{Lyon2017PRL, Sichau2019}, $\Delta\rho_{xx} = \rho_{xx}^{\nu} - \rho_{xx}^{\mathrm{dark}}$, requiring the subtraction of a measurement without microwaves. Figure \ref{fig3} (a) illustrates this analysis using an exemplary data set measured on our reference sample A for $n\approx$ 2 $\times$ 10$^{11}$ cm$^{-2}$. The resistivity under microwave radiation decreases on average by $\delta\Delta\rho_{xx}\simeq 0.55 {\rm{k}}\Omega$, implying an effective carrier density increase of $\delta n/n \simeq \delta \Delta \rho_{xx}/\bar{\Delta \rho_{xx}}\simeq 0.033$, where $\bar{\Delta\rho_{xx}}$ is the average longitudinal resistivity.
$\Delta\rho_{xx}$ reveals strong resonance peaks highlighted in red axially symmetric with respect to $B$ = 0. These resonances represent the resistive response to resonant spin-flips when h$\nu$ matches the Zeeman energy of the electrons in graphene. The peak centered a $B$ = 0 result from thermal activation in the weak localization regime, and weak peaks near |$B$| = 0.5 T are related to the intrinsic spin-orbit coupling band gap in graphene\cite{Sichau2019}. In the following, we focus our analysis on the prominent Zeeman peaks [additional data are available in the Supplementary Material].

The resonance line shapes and amplitudes encode information on electron spin diffusion and spin polarization. From a Lorentzian fit of the  resonance peaks, we obtain its \textbf{line width} $\Delta B_{\mathrm{res}}$, which is inversely proportional to the spin relaxation time\cite{Mani2012},

\begin{equation}
     \tau_\mathrm{S} = \frac{h}{4\pi \Delta E_{\mathrm{res}}} = \frac{h}{4\pi\cdot g\cdot \mu_B\cdot\Delta B_\mathrm{res}}.
\end{equation}

Here, $g$ is the electron $g$-factor in graphene \cite{Mani2012, Sichau2019}. The corresponding spin diffusion length ($\lambda_S$) is given by 
 \begin{equation}
     \lambda_\mathrm{S} = \sqrt{D\tau_S}\propto~\Delta B_\mathrm{res}^{-0.5}
\end{equation}
with the spin diffusion constant $D$ = 0.5$\cdot v_F\cdot l_e$ and the Fermi velocity\cite{Dias2017} $v_F = 10^6$ m/s. 
The average spin diffusion length in sample A, i.e., our unaltered CVD graphene, is (346$\pm$60) nm.

The \textbf{amplitude} of the microwave-induced resistance peak, on the other hand, is a measure of the transition probability between two spin states. For resistive detection of spin resonances, a sufficient number of electrons must be resonantly excited from the spin ground state and flip their spins \cite{Stein1983, Dobers1987, Dobers1988}. This probability is dictated by the number of available initial and final states, or more precisely, by the spin polarization that maps the difference between the number of spin-down $N_\downarrow$ and spin-up $N_\uparrow$ electrons. 

The open circles in Fig. \ref{fig3} (b) show the evolution of the resonance peak height with carrier concentration and carrier type for sample A (for a constant magnetic field, i.e., for a constant resonance frequency of 23 GHz) that is controlled by the gate voltage. The black solid line shows the charge neutrality point (CNP) in the absolute longitudinal sample resistance at $B$ = 0 T. The resonance amplitudes are largest around the CNP, where they culminate at 150 Ohms for this particular microwave frequency and power. As the Fermi level shifts away from the CNP towards higher densities, the amplitude rapidly decays and eventually vanishes as both bands become close to fully occupied. At a hole concentration of $p\approx$ 1 $\times$10$^{12}$ cm$^{-2}$, the peak height has dropped to approximately 10 $\Omega$. The red dashed line in Fig. \ref{fig3} (b) is a calculation of the spin polarization  using Eq. (\ref{eqSigma}) with $\alpha_s \simeq 650~\Omega$. This calculation reproduces reasonably well the overall exponential decay of the resonance amplitude with carrier concentration. At high electron and hole densities, the signal disappears in the resistive background. Both bands become close to fully occupied, rendering $\Sigma(n)$ of Eq. (1) zero, and other thermal carriers mask the resonant absorption processes. We note that the signal is proportional to  $\delta\Delta\rho_{xx}/\Delta\rho_{xx} = 
\delta \sigma_{xx}/\bar{\sigma}_{xx}$, with $\sigma_{xx}$ being the average conductivity enhancement by microwaves. At the CNP, i.e., where the signal is largest, $\bar{\sigma}_{xx}$ becomes minimal and $\delta \sigma_{xx}$ maximal.

\subsection{Nanomagnetic polarization boost}

\begin{figure}
   \includegraphics[width=0.35\textwidth]{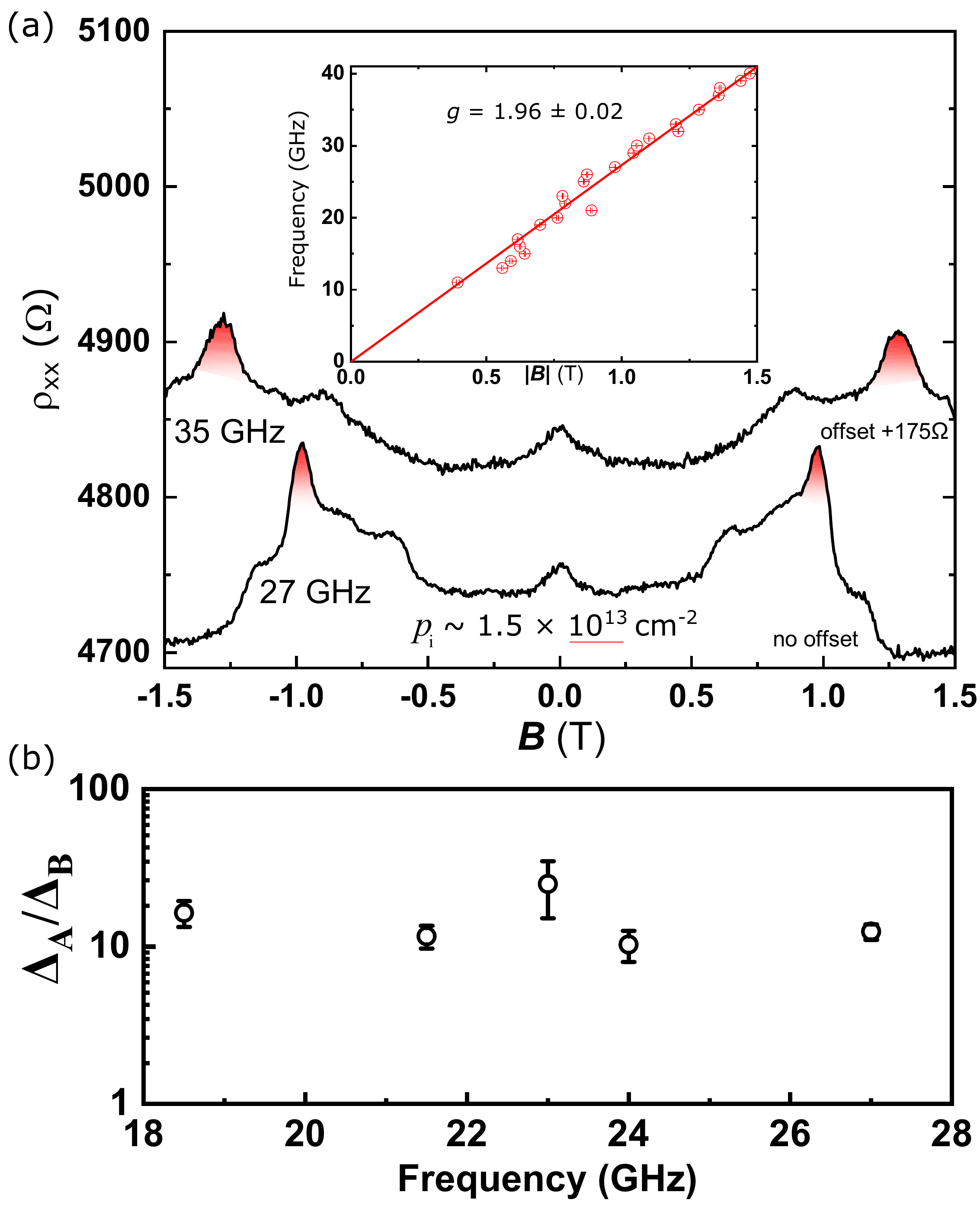}
  \caption{Sample B. (a) Absolute $\rho_{xx}$ shown for two exemplary frequencies, $\nu$ = 27 GHz and 35 GHz using a power of +21 dBm, shows strong Zeeman resonance peaks (red highlighted areas) and transitions originating from the existence of an intrinsic band gap (unmarked). Inset: Frequencies vs. resonance positions obtained from Lorentzian fits of the resonances \cite{Sichau2019}. The slope corresponds to a $g$-factor of 1.96$\pm$0.02. All measurements were performed with the (intrinsic) hole density of $p_i\approx$ 1.5 $\times$ 10$^{13}$ cm$^{-2}$, which is two orders of magnitude larger as for the data shown in Fig. \ref{fig3}(b). (b) Ratio of peak heights of $\Delta_\mathrm{A}$ (sample A at $n$ = 2$\times$10$^{11}$cm$^{-2}$) and $\Delta_\mathrm{B}$ (sample B at $p_i$) vs. frequency.} 
  \label{fig4}
\end{figure}

By placing a sheet of our monolayer CVD graphene over an artificial lattice of magnetic nanodots (sample B), we introduce a surplus of spin-polarized carriers in the vicinity of each nanodot, which amplifies the overall resistive response of the graphene layer under spin resonance. The platinum in the capping layer of the nanodots possesses a very high electron affinity, which induces strong $p$-doping and a high intrinsic hole concentration in the graphene of sample B ($p\approx$ 1.5 $\times$ 10$^{13}$ cm$^{-2}$). Figure \ref{fig4} (a) shows two typical measurements of the (absolute) longitudinal resistivity $\rho_{xx}$ on sample B under constant microwave radiation of frequencies $\nu$ = 27 GHz and 35 GHz. Despite the very high intrinsic hole concentration, which is two orders of magnitude larger than for the data shown in Fig.\ref{fig3} (and one order of magnitude larger than those densities for which the resonance peak had vanished in the resistive background), large resonances are observed directly in the resistivity. We stress that a similar density-dependent measurement as shown in  Fig. \ref{fig3}(b) was not feasible since the nanodots appear to screen the electric field from the back gate; even excessively large voltages only marginally affected the carrier density. In the following section, we will elaborate on a hypothetical density dependence. Based on Eq. (1) and the behavior seen in Fig. \ref{fig3}(b), we can expect an enhancement of the resonance amplitudes in sample B  at lower carrier concentrations. 

We evaluated the ESR frequency-dependence and plotted $\nu$ versus the occurrence of the Zeeman resonance in $B$ [inset Fig. \ref{fig4}(a)]. Additional data are available in the Supplementary Material. From a linear fit of the resulting dispersion, we can deduce the electron $g$-factor of 1.96$\pm$0.02, which confirms previous reports of ESR on graphene\cite{Mani2012, Sichau2019}. By comparing the resonance peak amplitudes of samples A and B, measured under the same experimental conditions (e.g., temperature, microwave power), we can stipulate a \textbf{relative} change in the spin polarization $N_\uparrow-N_\downarrow$ that is induced by the magnetic nanodots. Figure \ref{fig4}(b) shows that the ratio of the resonance amplitudes $\Delta_\mathrm{B}$ [measured in sample B at its intrinsic density] and $\Delta_\mathrm{A}$ [measured in sample A tuned to $n$ = 2$\times$10$^{11}$cm$^{-2}$] at the same frequencies is constant with a mean value of 15.2$\pm$4.4. By direct comparison the resonance peaks in sample A are still larger by a factor of 15, however, the carrier concentration is also 100$\times$ smaller.  

\subsection{Interpretation\label{sec:interpretation}}

\begin{figure}[!b]
   \includegraphics[width=0.45\textwidth, angle=0]{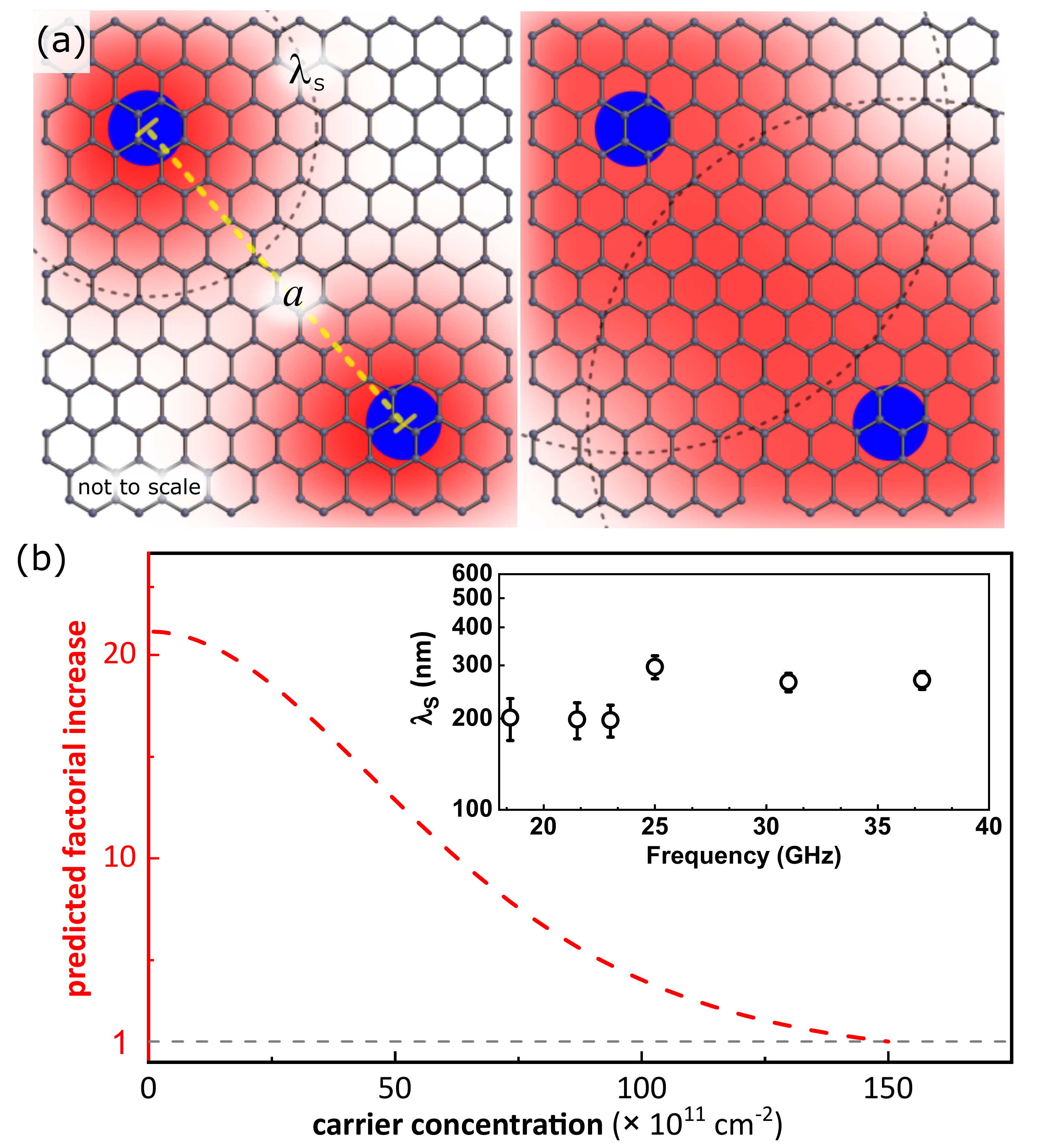}
   \caption{(a) Simplified schematic illustration of two contrasting diffusive regimes. Blue filled circles represent the nanodots, the red semitransparent areas [enclosed by a dashed circle for clarity] represent the spin diffusion length $\lambda_\mathrm{S}$, $a$ is the interdot distance. For $\lambda_\mathrm{S}/a < $1 (left panel), no excess polarization can build up. For  $\lambda_\mathrm{S}/a \geqslant$1 (right panel), a net excess spin polarization is maintained during the transit to the next 'spin hot spot', where the polarization is refreshed.  Parameter $a$ represents the mean interdot distance. (b) The predicted enhancement factor of the resonance amplitude for lower carrier concentrations. Inset: exemplary data points for $\lambda_\mathrm{S}$ versus $\nu$. } 
  \label{fig5}
\end{figure}

The confirmation of a $g$-factor of 1.96$\pm$0.02 in electron spin resonances experiments demonstrates that our measurements are probing the graphene layer and not the Pt/Co dots. The nanodots, however, enhance the ESR transition probability in the graphene by adding out of thermal equilibrium, spin-polarized carriers. We propose that the nanodots act as 'spin hot spots' that boost the polarization by carrier exchange, i.e., electrons enter the magnetic nanodots where they become spin-polarized before they are re-injected into the graphene. Carriers originating from the Fermi surface of cobalt have a spin polarization of about 20$\%$\cite{Busch1971, Kisker1982}. During the passage through the platinum, some of this polarization will be lost due to the small spin diffusion length in Pt\cite{Zhang2013}. The polarized carriers emerging from such a spin hot spot will be subject to additional sources for spin relaxation in the graphene. However, the spin diffusion length $\lambda_\mathrm{S}$ = (220$\pm$49) nm in sample B [inset Fig. \ref{fig5}(b)] is comparable to the distance between spin hot spots, enabling the conservation of an excess polarization before the electrons reach the next dot. These repeated processes across hundreds of spin hot spots between the voltage probes of the graphene lead to a higher (non-equilibrium) net spin polarization which we detect in the longitudinal resistance.

Figure \ref{fig5}(a) schematically illustrates the underlying principle using two dots for two distinct regimes in which the spin diffusion $\lambda_\mathrm{S}$ is much larger and smaller than the interdot distance $a$. The red highlighted areas that are surrounded by a dashed circle represent the average range electrons can propagate before the spin diffuses. The average spin diffusion length in sample B [inset Fig. \ref{fig5}(b)] is smaller than in sample A due to the different $l_\mathrm{e}$ that determines the spin diffusion constant and consequently $\lambda_\mathrm{S}$.

We note that each nanodot has a maximal stray field of approximately 59 mT, the total mean stray field of all dots is less than 1 mT averaged over the sample area, however. The stray field is thus too small to generate a detectable deviation from the previously reported $g$-factor of 1.95$\pm$0.01 or to  play any role in the polarization enhancement. 

The geometry of the lattice plays a menial role when the spin diffusion length exceeds the lattice constant. 
The model in section \ref{sec:model} is indeed based on the statistics that dictates the polarization enhancement, rendering geometrical details unimportant. A too closely packed lattice, however, would add too many carriers, essentially turning graphene into a metal with the benefit of the additional polarized carriers being nullified by the large carrier concentration. A sparse nanodot superlattice with $a\gg\lambda_s$, on the other hand, would not show any polarization enhancement signatures [left panel of Fig. \ref{fig5} (a)]. We also note that the morphology of free-standing graphene monolayers is subject to intrinsic rippling with a 1-nm-high corrugation normal to the surface\cite{Meyer2007, Geringer2009}. Graphene that was synthesized by CVD on copper foil and then transferred to a substrate has the tendency to form folds with heights of $<$ 10 nm. Graphene placed on a lattice of 10-nm-high nanodots will certainly be subject to additional corrugation, however, this artificial rippling and its induced strain is too weak to account for the polarization effect we observed. 

Before we conclude, we briefly revisit a hypothetical carrier concentration dependence of sample B. To estimate a potential change in the resonance peak height at low carrier concentrations, we reverse-engineered our model for the resonance amplitudes using the results at $p_i$ that correspond to a certain surplus spin polarization. We employ a model similar to Eq. (\ref{eqSigma}), with enhanced spin polarization by a (pessimistic) factor 5 with respect to the thermal polarization, where the Fermi functions are adjusted to reproduce this polarization enhancement (see Supplementary Material). As Fig. \ref{fig5} (b) shows, close to the CNP at a density of $\approx$ (1 - 2) $\times$ 10$^{11}$ cm$^{-2}$, the predicted resonance amplitudes should be a factor of 23$\times$ larger than at $p_i$.

\section{CONCLUSION}

In this study we have shown that the electron spin polarization of graphene can be boosted by placing monolayer CVD graphene across a substrate decorated with magnetic nanodots. Through carrier exchange, the magnetic impurities add a surplus of spin-polarized carriers. We emphasize that our system does not constitute a spin-valve device or a spintronics application in the classical sense that injects and detects a well-defined spin configuration. Here, we only outlined a possible route to amplify but also attenuate and control the electron spin polarization over macroscopic distances in graphene. We verified the enhanced spin polarization through resistively-detected electron spin resonance measurements, a method that probes the spin system as a whole. Utilizing a doping scheme for signal amplification has shown to be vital in optical signal transmission in fibers (i.e., erbium-doped fiber amplifiers), for example, and might also proof to be useful in van-der-Waals devices.


\begin{acknowledgments}
We want to acknowledge Andreas Stierle and Thomas F. Keller (German Electron Synchrotron, DESY) for providing SEM images and an EDX analysis of our samples. We also wish to acknowledge support by the Deutsche Forschungsgemeinschaft (DFG) within the Excellence Cluster Advanced Imaging of Matter (AIM, EXC 2056). Andreas Meyer from the department of chemistry of the Universit\"at Hamburg provided micelles for the deposition of the SiO$_2$ particles. All measurements were performed with \textit{nanomeas}\cite{nanomeas}. 
\end{acknowledgments}


\nocite{*}
\bibliography{manuscript_dots_arxiv.bbl}
\bibliographystyle{ieeetr}

\end{document}